\documentstyle[prb,preprint,aps]{revtex}

\begin{document}
\draft

\title{A differential cluster variation method for analysis of spiniodal decomposition in alloys}
\author{Zhi-Rong Liu$^1$ and Huajian Gao$^{1,2}$}
\address{$^1$ Max Planck Institute for Metals Research, D-70569 Stuttgart,
Germany}
\address{$^2$ Division of Mechanics and Computation, Department of Mechanical Engineering,
Stanford University, Stanford, California 94305, USA}

\maketitle

\begin{abstract}
A differential cluster variation method (DCVM) is proposed
for analysis of spinoidal decomposition in alloys.  In this method, lattice symmetry operations
in the presence of an infinitesimal composition gradient are utilized to
deduce the connection equations for the correlation functions and to
reduce the number of independent variables in the cluster
variation analysis.
 Application of the method is made to calculate
the gradient energy coefficient in the Cahn-Hilliard free energy
function and the fastest growing wavelength for spinodal decomposition
in Al-Li alloys. It is
shown that the gradient coefficient of congruently ordered Al-Li
alloys is much larger than that of the disordered system. In such
an alloy system, the calculated fastest growing wavelength is approximately
10 nm, which is an order of magnitude
larger than the experimentally observed domain size. This may provide a
theoretical explanation why spinodal
decomposition after a congruent ordering is dominated by the
antiphase boundaries.

\end{abstract}

\pacs{PACS: 64.75.+g, 81.30.-t, 05.70.Ln, 64.60.Cn}

\vspace{2mm}


\section{Introduction}

Phase transformation is critically important to the development of new
materials since it yields abundant microstructures in the
micro/mesoscopic scales. In general, there are two kinds of
kinetics for phase transformation, i.e., a nucleation process from
an initial metastable state or a spinodal decomposition
from an initial
unstable state. In the latter case, the phase transformation is
determined mainly by diffusion since there is no thermodynamic
barrier.\cite{1,2,3} Compared with that in a nucleation process,
the domain size distribution width in a spinodal decomposition
of binary systems is greatly compressed, which may produce rather
interesting nanostructures.\cite{4} Spinodal decomposition
has been commonly used to control grain structure in materials
such as Alnico alloys and Al-Li alloys to enhance the properties
of these materials.\cite{5,6,7}

The mostly widely used continuum theory to describe spinodal
decomposition was
presented by Cahn and Hilliard.\cite{8,9} The free energy of a
compositionally non-uniform alloy is expressed in the theory as a
Ginzburg-Landau expression:
\begin{equation}
F(\{c({\bf r})\})=\int[f_0(c)+\kappa(\nabla c)^2]d^3{\bf r},
\end{equation}
where $f_0(c)$ is the local free energy density of the homogeneous
system and $\kappa$ is the gradient energy coefficient. In the
initial stage of spinodal decomposition, the Cahn-Hilliard
equation can be approximately linearized and analytically solved.
When possible strain effect is ignored, one finds a fastest growing
wavelength\cite{9a}
\begin{equation}
\lambda_m=4\pi \left[ \kappa\left/\frac{\partial^2f_0(c)}{\partial
c^2}
                \right] ^{1/2}. \right.
\end{equation}
It has been recognized that $\lambda_m$ is an important quantity
to characterize domain size distribution in the decomposition process, hence could be
useful in
the design and evaluation of new materials. The local free energy density
$f_0(c)$, which describes the equilibrium thermal properties of
system, can be obtained from the highly accurate
calculation-of-phase-diagram (CALPHAD) method or other theoretical
methods such as cluster variation method (CVM) and molecular
dynamics (MD). However, the gradient energy coefficient $\kappa$,
as a parameter related to the non-equilibrium kinetics, has seldom been
directly calculated. Considering only the nearest-neighbor
interactions, $\kappa$ of an AB binary alloy with simple cubic lattice was expressed
under a point mean-field approximation (regular solution model)
as:\cite{8}
\begin{equation}
\kappa=\frac{2}{3}h_{0.5}^M r_0^2,
\end{equation}
in which $h_{0.5}^M$ is the heat of mixing per unit volume at the
composition $c=0.5$ and $r_0$ is the nearest-neighbor distance.
Recently, by using a supercell CVM approach, Asta and Hoyt
calculated the gradient coefficient of Ag-Al alloys and revealed
some behaviors of $\kappa$ that are different from the prediction
of the regular solution model.\cite{10} As these authors
pointed out, the values of $\kappa$ may have been influenced
by the
choice of parameters in their method because of the effects of
higher order terms in the gradient expansion of free energy.
In this paper, we present a differential cluster
variation method (DCVM) to calculate the gradient energy coefficient.
As will be shown, this method naturally avoids the influence of the higher
order terms in free energy expansion.

In phase transformation, the decomposition process can be
accompanied by an ordering process. Due to
the long-range diffusion associated with decomposition
and the
short-range diffusion associated with ordering, a congruent
ordering process frequently
occurs prior to the decomposition process.\cite{11} For example,
in Al-Li alloys, an important aerospace material, ordering
and decomposition processes result in precipitates of an
ordered $\delta'$ (Al$_3$Li) phase. If an congruent ordering
process takes place first, spinodal decomposition
does not occur from an initial disordered phase, but from a congruent
ordered phase, in which case Eq.(3) is no longer
applicable. It would be
interesting to explore the influence of congruent ordering on
the values of $\kappa$. By using the  proposed DCVM method, we will
calculate the gradient coefficient and the fastest
growing wavelength of Al-Li alloys, and discuss implications on
the actual decomposition process.

\section{Methods}

Cluster variation method (CVM) is a highly efficient microscopic
method to evaluate the configurational free energy and determine
the order-disorder phase transformation.\cite{12,13,14} In an
alloy system, the configurational free energy is determined by the
probabilities of all possible configurations. The essence of CVM
lies in using the probability of some small clusters to
approximately evaluate the probability of the entire system in
order to greatly decrease the number of independent variables. For
example, in a point approximation, one needs only to determine the
composition at every site such that the number of degrees of
freedom of the system decreases from $2^N$ to $N$ where $N$ is the
number of atoms in the system. For a homogeneous structure,
clusters related by the symmetry operations of the lattice can be
considered as identical, and the number of independent variables
can be further reduced. For example, for the schematic system
shown in Fig. 1 under the point approximation, the symmetry of the
lattice has the following relations:
\begin{eqnarray}
\nonumber x_1=x_3=\cdots=x_{2i+1}=\cdots; \\
x_2=x_4=\cdots=x_{2i}=\cdots,
\end{eqnarray}
where $x_i$ denotes the point probability at site $i$. By using
the above symmetry relations, the number of independent variables
decreases from $N$ to $n$ ($n=2$ in this case). In this way, the
equilibrium properties of the system, including the local free
energy density $f_0(c)$ in Eq. (1), can be easily determined.

When there are compositional fluctuations in the system,
the free energy can be expanded upon the homogeneous state.
In principle, the compositional fluctuations can be expressed in a
sum of Fourier components and the free energy expansion of any
Fourier component can be independently solved by the general
${\bf k}$-space formalism of standard fluctuation theory.\cite{14a}
In this paper, however, we directly consider a composition gradient
in the system (see Fig. 1) in order to evaluate the gradient coefficient
$\kappa$ in Eq. (1).
The introduction of the gradient will destroy the symmetry of
the lattice (point group and translations), and Eq. (4) is no
longer valid. The number of independent variables would become
much larger than that in the homogeneous case. One method to solve
this problem is to enlarge the number of independent variables,
i.e, to use a supercell, to include a non-uniform composition
variance. In the work of Asta and Hoyt, 98 planes were used to
calculate the gradient coefficient of Ag-Li alloys.\cite{10}

In fact, Eq. (1) is defined under the condition of small
composition gradient $\nabla c$. If $\nabla c$ is too large,
higher order terms in the free energy expansion may be
important and can not be simply ignored. To avoid the
influence of higher order terms, one can examine the response of
the system to an infinitesimal composition gradient. Such an
analysis could bring a great advantage: clusters related by
lattice symmetry operations under the uniform state can be
related by some equations under an infinitesimal gradient
variance, and one needs not to consider a supercell. We
develop such a method in the following.

Assume there are two kinds of infinitesimal variances applied upon
the system: (1) an infinitesimal composition gradient $\nabla
c=d{\bf g}$ which keeps the composition at the origin as $c$ and
(2) an infinitesimal uniform composition variance $dc$. In general,
the variation of the correlation function of a cluster can be
expanded as (we use the correlation function to effectively
replace the probability function of the cluster in order to
simplify the statement):
\begin{equation}
\xi(c,d{\bf g},dc)=\xi(c)+\left(\frac{\partial\xi}{\partial{\bf
g}}\cdot d{\bf g} +\frac{\partial\xi}{\partial c}dc \right)
+\frac{1}{2} \left[\frac{\partial^2\xi}{\partial{\bf
g}\partial{\bf g}}: d{\bf g}d{\bf g}
+2\frac{\partial^2\xi}{\partial{\bf g}\partial c}\cdot d{\bf g}dc
+\frac{\partial^2\xi}{\partial c^2}(dc)^2 \right],
\end{equation}
where $\frac{\partial\xi}{\partial{\bf g}},
\frac{\partial\xi}{\partial c}, \frac{\partial^2\xi}{\partial{\bf
g}\partial{\bf g}}, \ldots$ are expansion coefficients to be determined
by minimizing the free energy. We have expanded $\xi$
only to second order terms because the gradient energy in
Eq. (1) is related to the second order terms of the free
energy. Now consider a translation operation (with a displacement
vector ${\bf L}$) which is a symmetry operation under the uniform
case. When there is a gradient variance $d{\bf g}$, the system
after the translation is not identical to that before the
translation. The composition of the system changed by $-{\bf L}\cdot
d{\bf g}$ after translation. However, if one adds a uniform
composition variance ${\bf L}\cdot d{\bf g}$ to it, the system
will be identical to that before the translation. Therefore, for two
clusters located at {\bf r} and {\bf r+L} which are related by the
translation operation, the symmetry property gives
\begin{equation}
\xi_{\bf r+L}(c,d{\bf g},dc)=\xi_{\bf r}(c, d{\bf g}, dc+{\bf
L}\cdot d{\bf g}).
\end{equation}
By substituting Eq. (5) into the above equation, one obtains
\begin{eqnarray}
  \nonumber \xi_{\bf r+L}(c) & + & \left(\frac{\partial\xi_{\bf
r+L}}{\partial{\bf g}}\cdot d{\bf g} +\frac{\partial\xi_{\bf
r+L}}{\partial c}dc \right) +\frac{1}{2}
\left[\frac{\partial^2\xi_{\bf r+L}}{\partial{\bf g}\partial{\bf
g}}: d{\bf g}d{\bf g} +2\frac{\partial^2\xi_{\bf
r+L}}{\partial{\bf g}\partial c}\cdot d{\bf g}dc
+\frac{\partial^2\xi_{\bf r+L}}{\partial c^2}(dc)^2
\right] \\
 &\lefteqn{  =   \xi_{\bf r}(c)+\left[\frac{\partial\xi_{\bf
r}}{\partial{\bf g}}\cdot d{\bf g} +\frac{\partial\xi_{\bf
r}}{\partial c}(dc+{\bf
L}\cdot d{\bf g}) \right]  } \nonumber\\
  && \lefteqn{ +\frac{1}{2}
\left[\frac{\partial^2\xi_{\bf r}}{\partial{\bf g}\partial{\bf
g}}: d{\bf g}d{\bf g} +2\frac{\partial^2\xi_{\bf r}}{\partial{\bf
g}\partial c}\cdot d{\bf g}(dc+{\bf L}\cdot d{\bf g})
+\frac{\partial^2\xi_{\bf r}}{\partial c^2}(dc+{\bf L}\cdot d{\bf
g})^2 \right], }
\end{eqnarray}
which gives the following connection equations for the expansion
components:
\begin{eqnarray}
\xi_{\bf r+L}(c)=\xi_{\bf r}(c),\\
\frac{\partial\xi_{\bf r+L}}{\partial{\bf
g}}=\frac{\partial\xi_{\bf r}}{\partial{\bf g}}
+\frac{\partial\xi_{\bf r}}{\partial c}{\bf L}, \\
\frac{\partial\xi_{\bf r+L}}{\partial c}=\frac{\partial\xi_{\bf
r}}{\partial c}, \\
 \frac{\partial^2\xi_{\bf
r+L}}{\partial{\bf g}\partial{\bf g}} =\frac{\partial^2\xi_{\bf
r}}{\partial{\bf g}\partial{\bf g}} +2\frac{\partial^2\xi_{\bf
r}}{\partial{\bf g}\partial c}{\bf L} +\frac{\partial^2\xi_{\bf
r}}{\partial c^2}{\bf L}{\bf L},\\
\nonumber \ldots
\end{eqnarray}
For a point group operation, similar equations can be
given. Based on these equations, for a group of clusters
related by the symmetry operations of the lattice, their
correlation functions (or probability functions) can be expressed
by that of a representative independent cluster even if there is
an (infinitesimal) gradient variance.

To deduce the equations to solve the expansion coefficients, we
expand the differential of the free energy as:
\begin{eqnarray}
\nonumber \lefteqn{   \frac{\partial F(c,d{\bf
g},dc)}{\partial\xi_i}
  =\frac{\partial F(c)}{\partial\xi_i}
  +\sum_{j}\frac{\partial^2 F(c)}{\partial \xi_i \partial \xi_j}d\xi_j
  +\frac{1}{2}\sum_{j,k}\frac{\partial^3 F(c)}{\partial \xi_i \partial \xi_j \partial \xi_k}d\xi_jd\xi_k } \\
\nonumber
  &=& \frac{\partial F(c)}{\partial\xi_i}
  +\sum_{j}\frac{\partial^2 F(c)}{\partial \xi_i \partial \xi_j}
  \left\{
  \left(\frac{\partial\xi_j}{\partial{\bf
g}}\cdot d{\bf g} +\frac{\partial\xi_j}{\partial c}dc \right)
+\frac{1}{2} \left[\frac{\partial^2\xi_j}{\partial{\bf
g}\partial{\bf g}}: d{\bf g}d{\bf g}
+2\frac{\partial^2\xi_j}{\partial{\bf g}\partial c}\cdot d{\bf
g}dc
+\frac{\partial^2\xi_j}{\partial c^2}(dc)^2 \right] \right\} \\
 && +\frac{1}{2}\sum_{j,k}\frac{\partial^3 F(c)}{\partial \xi_i
\partial \xi_j \partial \xi_k}
\left(\frac{\partial\xi_j}{\partial{\bf g}}\cdot d{\bf g}
+\frac{\partial\xi_j}{\partial c}dc \right)
\left(\frac{\partial\xi_k}{\partial{\bf g}}\cdot d{\bf g}
+\frac{\partial\xi_k}{\partial c}dc \right).
\end{eqnarray}
For a non-point cluster, the minimization of the free energy gives
\begin{equation}
\frac{\partial F(c,d{\bf g},dc)}{\partial\xi_i}=0.
\end{equation}
For a point cluster, the free energy should be minimized under the
composition constraint as
\begin{equation}
\frac{\partial F(c,d{\bf g},dc)}{\partial\xi_i}=\mu_i,
\end{equation}
where $\mu_i$ is the chemical potential at site $i$
which controls the composition in the system. It has a
uniform value at different sites in a homogeneous
system.\cite{14} When there is a gradient composition variance,
$\mu_i$ also varies with gradient. Combining Eqns. (12-14), the
equations to determine the expansion coefficients,
$\frac{\partial\xi}{\partial{\bf g}}, \frac{\partial\xi}{\partial
c}, \frac{\partial^2\xi}{\partial{\bf g}\partial{\bf g}}, \ldots$,
can be easily obtained.

After solving the expansion coefficients, one can calculate the
free energy of the system under any infinitesimal gradient and
uniform variances as:
\begin{equation}
F(c,d{\bf g},dc)
  =F(c) + \sum_{i}\frac{\partial F(c)}{\partial \xi_i}d\xi_i
  + \frac{1}{2}\sum_{i,j}\frac{\partial^2 F(c)}{\partial \xi_i \partial \xi_j}d\xi_i d\xi_j
\end{equation}
The gradient energy coefficient $\kappa$ is then calculated
according to Eq. (1). Since the free energy is expressed as an
energy part and an entropy part in CVM, one can further calculate
the contributions of the energy and the entropy to the gradient
coefficient respectively. In the regular solution model, the
entropy depends only on the point probability, hence it has no
contribution to the gradient coefficient.\cite{8} When the
correlations between different points are considered (such as
CVM), as we will demonstrate next, the entropy will
give contribution to the value of $\kappa$. Eq. (15) can also be
used to evaluate $\partial^2f_0(c)/\partial c^2$, a key quantity
for evaluating the fastest growing wavelength [see Eq.
(2)], since it is related to the second expansion term of the free
energy under a uniform composition variance.

The method proposed here can be generalized to including higher
order terms in the expansion of the free energy and other kinds of
infinitesimal variance. This will not be pursued in this
paper.

\section{Results and Discussions}

Now we apply the above method to
calculate the gradient energy coefficient along the (100)
direction in the f.c.c. lattice.

To compare the new method with the regular solution
approximation\cite{8}, we first consider an exemplary AB spinodal
system with the nearest-neighbor interaction. The value of
the nearest-neighbor effective pair interaction is arbitrarily
chosen as $-100k_B$ ($k_B$ is the Boltzmann constant). The nearest
neighbor pair is used as the basic cluster in CVM calculation for
this case. This is a very ``pure" spinodal system that experiences a
decomposition into a mixture of A-rich and B-rich disordered
phases at low temperatures. There is no other ordering or
decomposition process in the system. The spinodal temperature at a
composition $c=0.5$ is determined as $T_c=1090K$ in our
calculation.

The calculation results of the gradient energy coefficient
$\kappa$ as functions of temperature $T$ are shown in Fig. 2 for
the uniform composition $c=0.5$ (solid) and $c=0.25$ (dashed). The
result of the regular solution model is also plotted in Fig. 2 as
the dotted line for comparison. For $c=0.5$, the calculated
$\kappa$ value is very close to the value predicted by the regular
solution model at high temperatures ($T>600K$). At low
temperatures, the $\kappa$ value rapidly increases with decreasing
temperature. When $c=0.25$, the situation is similar, except that
$\kappa$ is roughly 4\% higher than the regular-solution-model
value at high temperatures.

The values of $\kappa$ coming from energy and entropy contribution
(denoted as $\kappa_E$ and $\kappa_S$) are calculated respectively
and the results are plotted in Fig. 3. It can be seen that the
entropy gives no important contribution to $\kappa$ at high
temperatures, which is consistent with the prediction of the
regular solution approximation. However, when the temperature is
low enough ($T<300K$), the entropy contributes an non-negligible
negative value to the gradient coefficient $\kappa$. This result
is comprehensible from the consideration that pair correlations
involved in the CVM calculation is important at low temperatures.

In Fig. 4 the calculated values of $\kappa$ are plotted as a
function of composition at temperatures $T=500$ and $700$ $K$.
Since the curves are symmetrical about $c=0.5$, only the parts in
the range $0<c<0.5$ are plotted. It can be seen that the $\kappa$
value at $c=0.5$ is the closest to the regular solution
approximation value. When the composition deviates from $0.5$,
$\kappa$ increases steadily. An increment of about 30\% and 50\%
is obtained for $T=500$ and $700$ $K$ respectively when the
composition decreases to $c=0$. Different from Ref. 11, no
abnormal behavior is observed at $c=0$ in Fig. 4.

From the above results, we see that the gradient coefficient
$\kappa$ increases with decreasing temperature and composition
(for $c<0.5$). These characteristics are consistent with previous
supercell CVM calculation results on Ag-Al alloys, where a similar
spinodal decomposition process on disordered phase was
considered.\cite{10} In comparison, our results are closer to the
regular solution approximation values, while the previous
supercell results for Ag-Al are about three to four times larger
than the mean-field values.

Then we turn to a realistic system, Al-Li alloys, one of the best
known examples for material ordering strengthening.\cite{6,7,15}
In this system, ordering and decomposing processes coexist. Due to
the difference in diffusion length scales, ordering process usually
occurs much faster than the decomposing process.\cite{16,17} We
first optimize the ordered parameter at every composition, i.e.,
use the conventional CVM process to determine the ordered state of
the homogeneous system. Then we use the DCVM method to
investigate the response of the system to a gradient variance. In
other words, we calculate the gradient coefficient of the
congruent ordered phase. The effective pair interaction parameters
are taken from the study of Garland et al,\cite{18} where the
nearest and second-nearest-neighbor effective pair interactions
were chosen as 210 $k_B$ and -105 $k_B$, respectively.

The calculated gradient energy coefficient $\kappa$, together with
the contributions from energy and entropy parts ($\kappa_E$ and
$\kappa_S$), are shown in Fig. 5 as functions of Li composition at
$T=500K$. An extraordinary characteristic of the result is that
the calculated $\kappa$ of Al-Li alloys here is about one order of
magnitude larger than that in Ag-Al and Al-Zn alloys.\cite{10,18a}
This indicates that the congruent ordering greatly affects the
gradient coefficient of the system. When a congruent ordering
occurs in Al-Li alloys, the face-center sites of f.c.c. lattice
are mainly occupied by Al atoms, resulting in a very low local Li
composition at these sites. This may be an important source of the
large $\kappa$ in Al-Li alloys if one remembers that $\kappa$
increases for small composition in disordered phase (Fig.4 and
Ref. 11).

Fig. 5 also indicates that the entropy contribution to the
gradient coefficient, $\kappa_S$, is positive in value. It even
exceeds the contribution from the energy. This reveals a strong
correlation effect between points at different sites for the
congruent ordered phase.

By using the $\kappa$ value in Fig. 5, the fastest growing
wavelength $\lambda_m$ is calculated according to Eq. (2) for
various Li compositions. It appears that $\lambda_m$ is on the
order of about 10 nm (Fig. 6). According to previous experiments
and microscopic simulations, the modulation wavelength of Al-Li
alloys in the early stage is about $1\sim2$ nm.\cite{19,20} The
discrepancy between the current calculated $\lambda_m$ value and
the experiments and the pervious simulation results may be
attributed to the existence of antiphase boundaries in the
congruent ordered phase. When a spinodal decomposition occurs
after a congruent ordering, the microscopic simulation reveals
that the structural evolution in the decomposition process is
dominantly governed by the antiphase boundaries produced in the
congruent ordering process, e.g., the equilibrium disordered phase
appears and grows at the antiphase boundaries.\cite{11,16} The
antiphase boundaries play the role of nuclei for phase
decomposition. Thus the modulation wavelength is determined by the
distribution of antiphase boundaries. However, inside the spinodal
region of the phase diagram, any fluctuation, no matter how small
in degree, decreases the free energy and the system is unstable.
So it is quite confusing why a spinodal decomposition does not
occur inside the congruent ordered domains even if such process
would be spontaneous. The calculation result of the fastest
growing wavelength in Fig. 6 provides a key to answer this
problem. For the spinodal decomposition process, there is a
critical wavelength $\lambda_c=\lambda_m/\sqrt{2}$. Fourier
components with modulation wavelength $\lambda<\lambda_c$ will
decay while those with $\lambda>\lambda_c$ will grow. According to
the results in Fig. 6, the critical wavelength $\lambda_c$ is
larger than the size of congruent ordered domains observed in
experiments and previous simulation. Therefore, the spinodal
process is suppressed inside the congruent ordered domains. On the
basis of this explanation, when a decomposition process occurs
after the congruent ordering, there is no essential difference for
the systems inside and outside the spinodal region since the
antiphase boundaries act as nuclei for decomposition. It also
implies that an attempt\cite{21} to distinguish if the Al-Li
alloys is spinodal or not after the congruent ordering is not
important to the actual process.

It should be noted that the strain energy is not considered in the
current calculation of the fastest growing wavelength $\lambda_m$.
If the strain effect is involved, $\lambda_m$ is expected to be larger
than the current values, which will not change the above conclusions.

Fig. 7 shows the gradient energy coefficient $\kappa$ of Al-Li
alloys as functions of temperature when the Li composition is
fixed as $15\%$. Different from the case of the disordered phase
(Fig. 3), $\kappa$ of Al-Li alloys increases with increasing
temperature. Only when the temperature is close to the stability
limit of the ordered phase does the $\kappa$ value drop rapidly.
Accordingly, the curve of the fastest growing wavelength exhibits
a peak near the limiting temperature (see the inserted graphics
in Fig. 7).


\section{Summaries}
In summary, a differential cluster variation method (DCVM) is
developed this paper and used to calculate the gradient energy coefficient
($\kappa$) in the Cahn-Hilliard course-grained free energy for
spinodal decomposition. The symmetry equations for clusters
related by symmetry operations of a lattice are deduced under
an infinitesimal composition gradient in the system. This treatment
greatly reduces the number of independent variables in comparison
with a supercell CVM analysis. The DCVM is an intrinsic method
to determine different order terms in the gradient expansion of
the free energy. The value of $\kappa$ of two systems with
f.c.c. lattice are calculated by this new method. It is shown
that the $\kappa$ values for the system with only the
nearest-neighbor interaction are very close to the prediction
of the point mean-field approximation (regular solution model)
for most system states, while non-negligible differences exist
at low temperatures and extreme compositions. For the Al-Li
alloys where a congruent
ordering occurs prior to the spinodal decomposition, $\kappa$ is
found to be much larger than the disordered system. The fastest
growing wavelength is calculated to be approximately 10 nm
at 500 $K$, which is one order of magnitude larger than the
experimentally observed domain size. This provides a theoretical
explanation to the previous discovery that the spinodal
decomposition after congruent ordering is dominated by the
antiphase boundaries produced in the congruent ordering
process.

\section*{Acknowledgment}

This work was supported by a Max Planck Post-Doc Fellowship for
ZL. The work is also supported by US National Science Foundation
through Grant CMS-0085569. We thank Prof. Long-Qing Chen at the
Pennsylvania State University for very helpful discussions.

\begin{figure}[tbp]
\caption{Schematic graphics of a one-dimensional lattice with
compositional gradient variance.}
\end{figure}

\begin{figure}[tbp]
\caption{The gradient energy coefficient $\kappa$ as function of
temperature $T$ (in units of $K$) for a disordered f.c.c. binary
system with the nearest-neighbor interaction -100 $k_B$; $\kappa$
is measured in units of $k_B/a_0$ where $a_0$ is the lattice
constant. The dotted line denotes the predicted value of the
regular solution model.}
\end{figure}

\begin{figure}[tbp]
\caption{The energy part ($\kappa_E$) and the entropy part
($\kappa_S$) of the gradient coefficient as function of
temperature (in units of $K$) for a disordered system with the
nearest-neighbor interaction -100 $k_B$. The composition is fixed
as $c=0.5$. $\kappa_E$ and $\kappa_S$ are measured in units of
$k_B/a_0$ where $a_0$ is the f.c.c. lattice constant. }
\end{figure}

\begin{figure}[tbp]
\caption{The gradient energy coefficient $\kappa$ (in units of
$k_B/a_0$) of a disordered f.c.c. binary system as function of
composition $c$. The dotted line denotes the predicted value of
the regular solution model.}
\end{figure}

\begin{figure}[tbp]
\caption{The gradient coefficient ($\kappa$) and its energy part
($\kappa_E$) and entropy part ($\kappa_S$) of the congruent
ordered Al-Li alloys as function of Li composition when he
temperature is fixed at $T=500K$.}
\end{figure}

\begin{figure}[tbp]
\caption{The fastest growing wavelength $\lambda_m$ of the
congruent ordered Al-Li alloys calculated according to Eq. (2).
The temperature is fixed at $T=500K$.}
\end{figure}

\begin{figure}[tbp]
\caption{The gradient coefficient of the congruent ordered Al-Li
alloys as function of temperature when the Li composition is
fixed at $15\%$. The inserted graphics is the corresponding
fastest growing wavelength.}
\end{figure}


\vspace{2mm}

\begin{references}
\bibitem{1} J. D. Gunton, M. San Miguel, and P. S. Sahni, in {\it
Phase Transitions and Critical Phenomena}, edited by C. Domb and
J. L. Lebowitz (Academic Press, London, 1983), Vol. 8, p. 267.

\bibitem{2} K. Binder, in {\it Materials Science and Technology},
edited by R. W. Cahn, P. Haasen, and E. J. Kramer (VCH, Weinheim,
1991), Vol. 5, p. 405.

\bibitem{3} A. J. Bray, Adv. Phys. {\bf 43}, 357 (1994).

\bibitem{4} S. Herminghaus, K. Jacobs, K. Mecke, J. Bischof, A. Fery, M.
Ibn-Elhaj, and S. Schlagowski, Science {\bf 282}, 916 (1998).

\bibitem{5} B. Cullity, {\it Introduction to magnetic materials}
(Addison-Wesley, Reading, MA, 1 edition, 1972).

\bibitem{6} J. M. Silcock, J. Inst. Metals {\bf 88}, 357
(1959-60).

\bibitem{7} A. J. McAlister, Bull. Alloy Phase Diag. {\bf 3}, 177
(1982).

\bibitem{8} J. W. Cahn, and J. E. Hilliard, J. Chem. Phys. {\bf
28}, 258 (1958).

\bibitem{9} J. W. Cahn, and J. E. Hilliard, J. Chem. Phys. {\bf
31}, 688 (1959).

\bibitem{9a} J. W. Cahn, Acta Mater. {\bf 9}, 795 (1961).

\bibitem{10} M. Asta and J. J. Hoyt, Acta Mater. {\bf 48}, 1089 (2000).

\bibitem{11} A. G. Khachaturyan, {\it Theory of Structural
Transformations in Solids} (Wiley, New York, 1983).

\bibitem{12} R. Kikuchi, Phys. Rev. {\bf 81}, 988 (1951).

\bibitem{13} R. Kikuchi, J. Chem. Phys. {\bf 60}, 1071 (1974).

\bibitem{14} J. M. Sanchez and D.de Fontaine, Phys. Rev. B {\bf
21}, 216 (1980).

\bibitem{14a} H. E. Cook and D. de Fontaine, Acta Metall. {\bf 17}, 915 (1969).

\bibitem{15} E. Nembach, Prog. Mater. Sci. {\bf 45}, 275 (2000).

\bibitem{16} L. Q. Chen and A. G. Khachaturyan, Acta Metall. Mater. {\bf 39},
2533 (1991).

\bibitem{17} Z. R. Liu, B. L. Gu, H. Gui, and X. W. Zhang, Phys.
Rev. B {\bf 59}, 16 (1999).

\bibitem{18} J. S. Garland and J. M. Sanchez, in {\it Kinetics of
Ordering Transformations in Metals}, edited by H. Chen and V. K.
Vasudevan, p207 (TMS, Warrendale, PA, 1992).

\bibitem{18a} J. Mainville, Y. S. Yang, K. R. Elder, M. Sutton, K. F. Ludwig, G. B.
Stephenson, Phys. Rev. Lett. {\bf 78}, 2787 (1997).

\bibitem{19} D. Z. Che, S. Spooner, and J. J. Hoyt, Acta
Mater. {\bf 45}, 1167 (1997).

\bibitem{20} R. Poduri and L. Q. Chen, Acta Mater. {\bf 46}, 3915
(1998).

\bibitem{21} S. Banerjee, A. Arya, and G. P. Das, Acta Mater. {\bf
45}, 601 (1997).

\end{references}
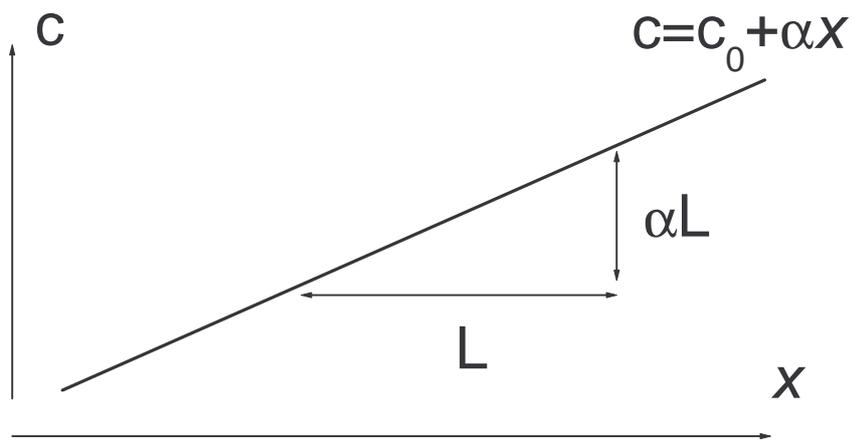
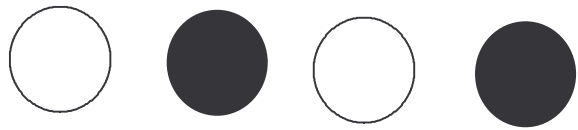
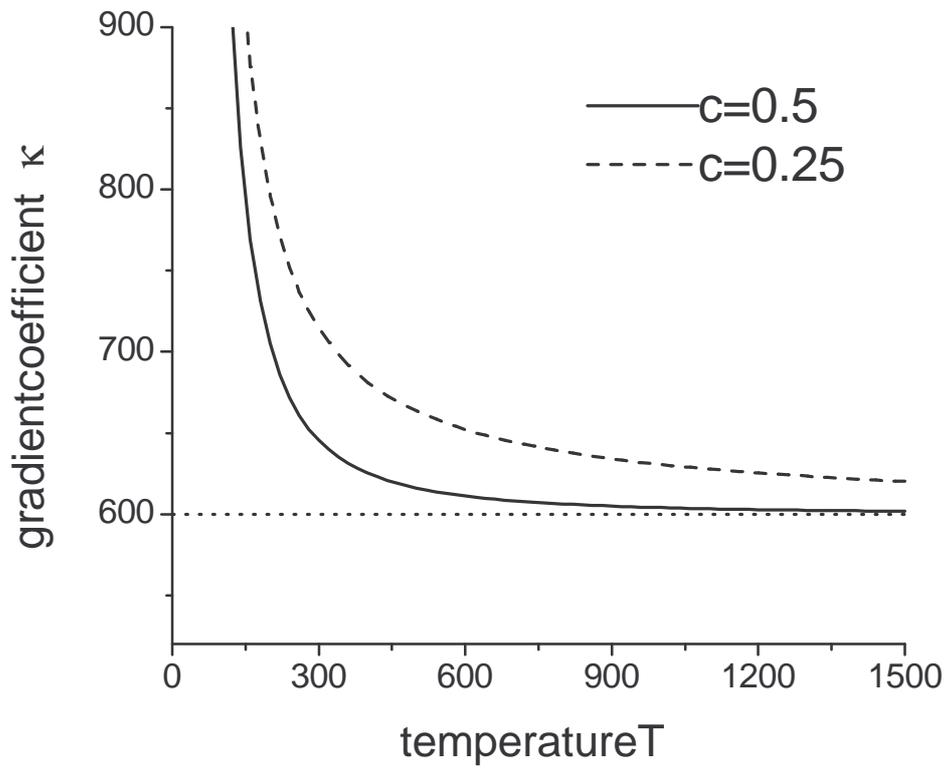
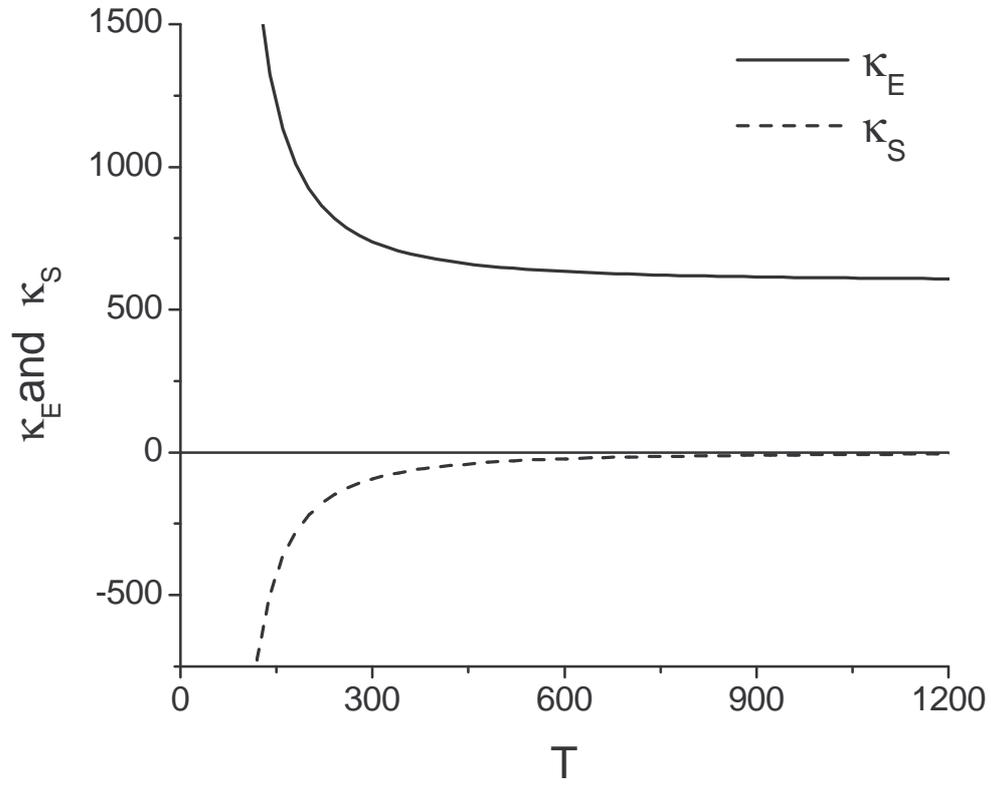
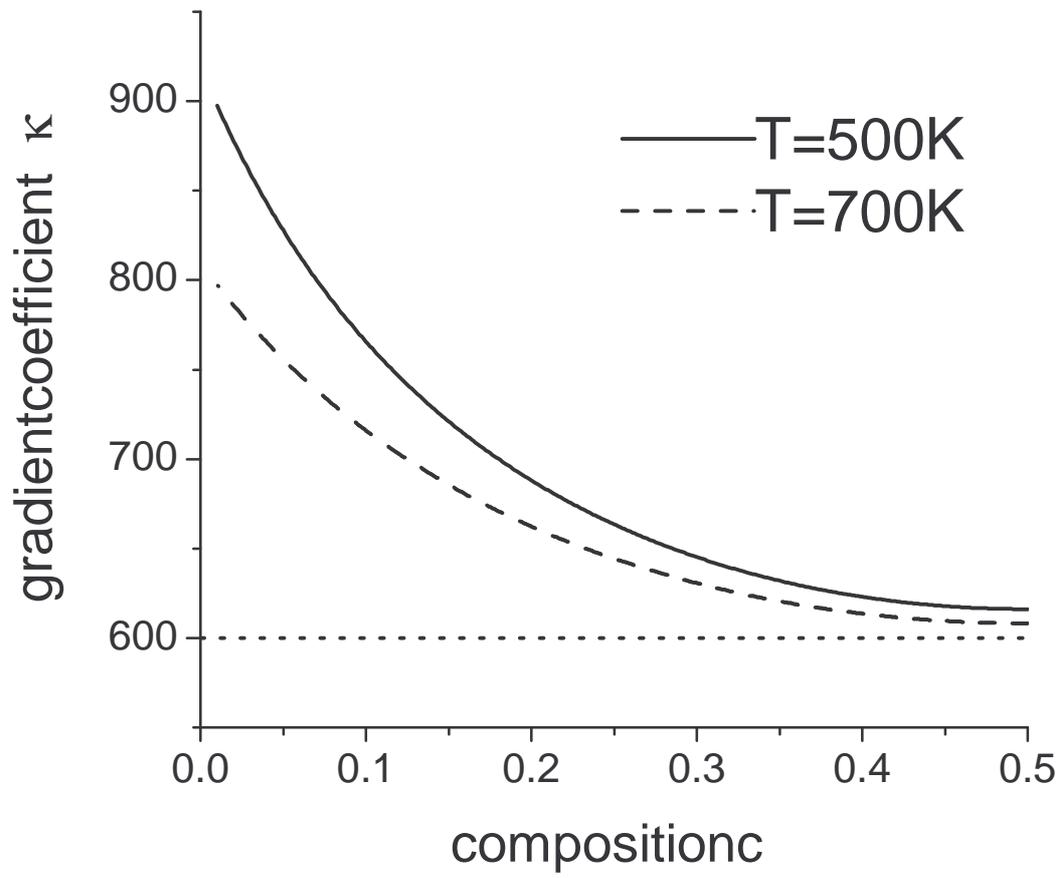
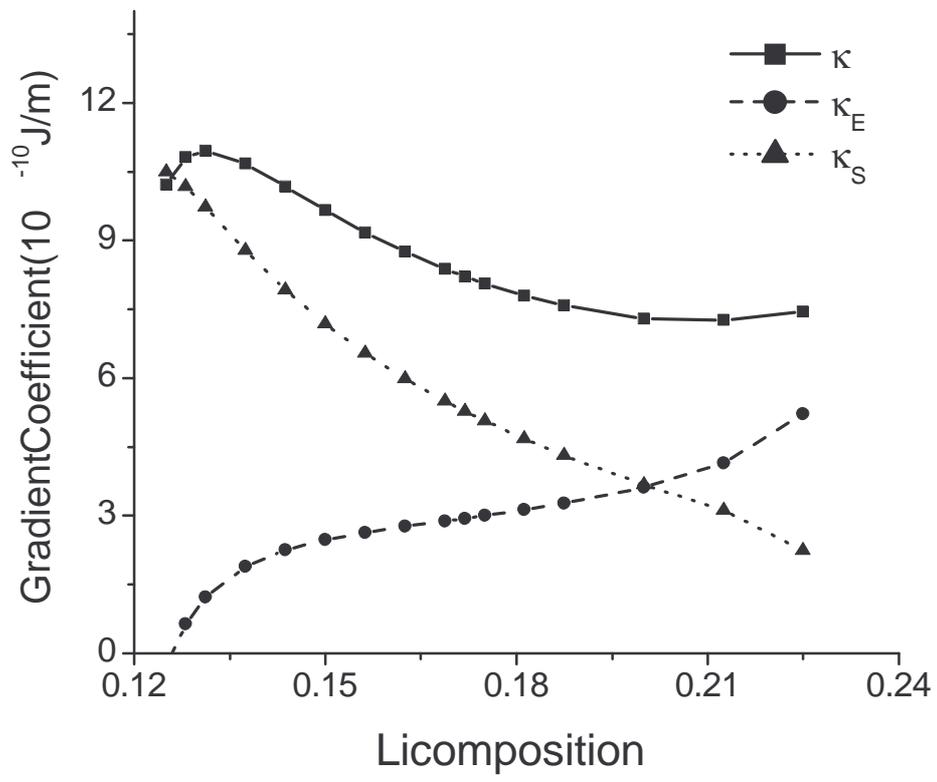
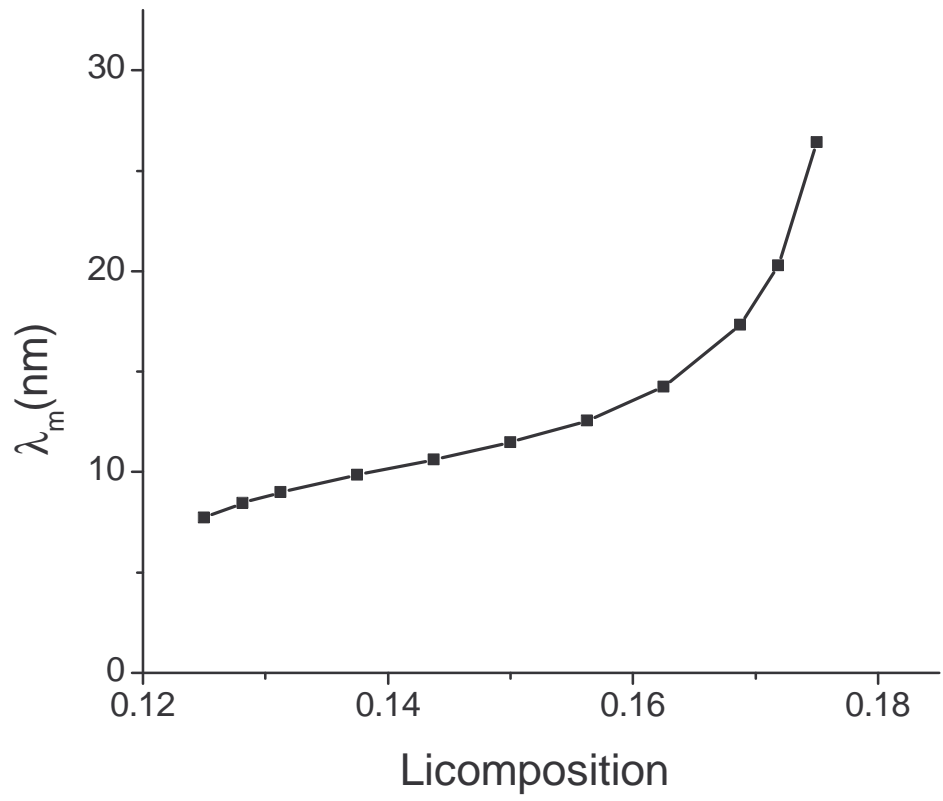
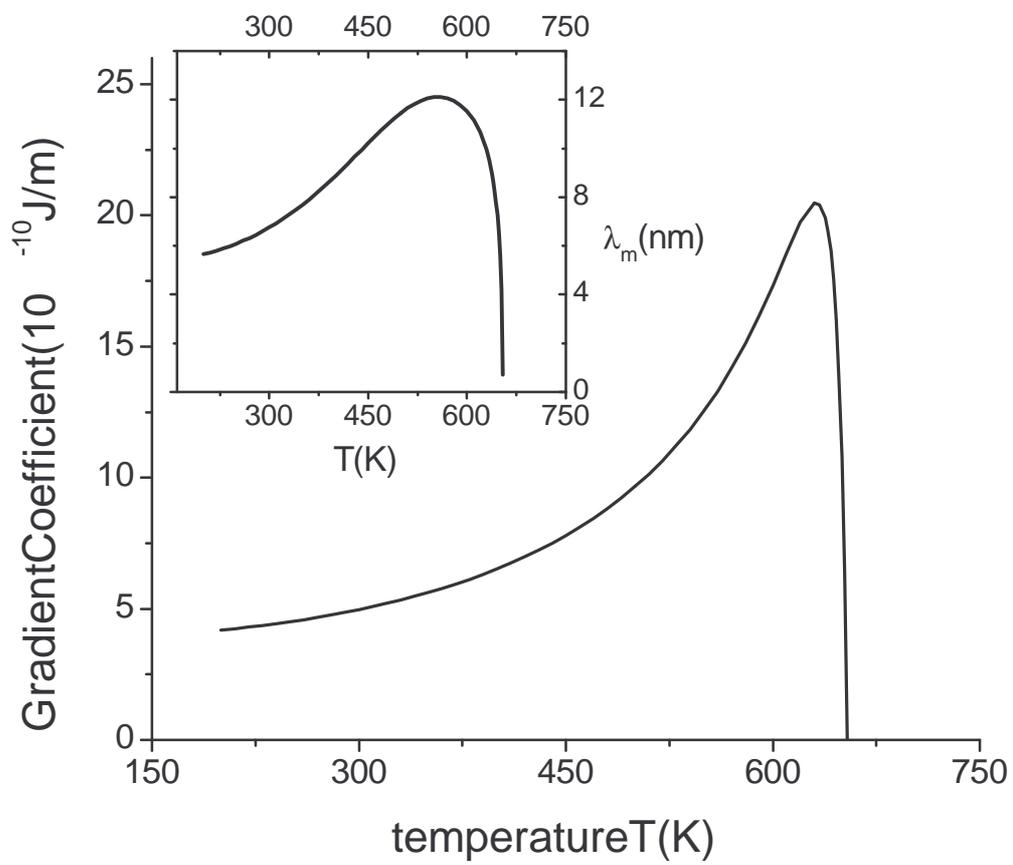
\end{document}